\documentclass[aps,pra,twocolumn,superscriptaddress,notitlepage,groupedaddress]{revtex4-2}

\usepackage{graphicx}
\usepackage{mathptmx, textcomp}
\usepackage[usenames]{xcolor}
\usepackage{dcolumn}
\usepackage{bm}
\usepackage{amssymb, amsmath}
\usepackage[colorlinks,urlcolor=blue,citecolor=blue,linkcolor=blue]{hyperref}
\usepackage{cleveref}
\usepackage{siunitx}
\usepackage{physics}
\usepackage{verbatim}
\usepackage{amsmath}
\usepackage[normalem]{ulem}
\usepackage{lineno}

\begin{document}

\title{Life and death of the Bose polaron}

\author{Magnus G. Skou$^1$}
\author{Kristian K. Nielsen$^{1,2}$}
\author{Thomas G. Skov$^1$}
\author{Andreas M. Morgen$^1$}
\author{Nils B. J{\o}rgensen$^1$}
\author{Arturo Camacho-Guardian$^{3}$}
\author{Thomas Pohl$^1$}
\author{Georg M. Bruun$^{1,4}$}
\author{Jan J. Arlt$^1$}
\affiliation{$^1$ Center for Complex Quantum Systems, Department of Physics and Astronomy, Aarhus University, Ny Munkegade 120, DK-8000 Aarhus C, Denmark.}
\affiliation{$^2$ Max-Planck Institute for Quantum Optics, Hans-Kopfermann-Str. 1, D-85748 Garching, Germany.}
\affiliation{$^3$ Departamento de F{\'i}sica Qu{\'i}mica, Instituto de F{\'i}sica, Universidad Nacional Aut{\'o}noma de M{\'e}xico, Apartado Postal 20-364, Ciudad de M{\'e}xico C.P. 01000, Mexico.}
\affiliation{$^4$ Shenzhen Institute for Quantum Science and Engineering and Department of Physics, Southern University of Science and Technology, Shenzhen 518055, China.}

\date{\today}

\begin{abstract}
Spectroscopic and interferometric measurements complement each other in extracting the fundamental properties of quantum many-body systems. While spectroscopy provides precise measurements of equilibrated energies, interferometry can elucidate the dynamical evolution of the system. For an impurity immersed in a bosonic medium, both are equally important for understanding the quasiparticle physics of the Bose polaron. Here, we compare the interferometric and spectroscopic timescales to the underlying dynamical regimes of the impurity dynamics and the polaron lifetime, highlighting the capability of the interferometric approach to clearly resolve polaron dynamics. In particular, interferometric measurements of the coherence amplitude at strong interactions reveal faster quantum dynamics at large repulsive interaction strengths than at unitarity. These observations are in excellent agreement with a short-time theoretical prediction including both the continuum and the attractive polaron branch. For longer times, qualitative agreement  with a many-body theoretical prediction which includes both branches is obtained. Moreover, the polaron energy is extracted from interferometric measurements of the observed phase velocity in agreement with previous spectroscopic results from weak to strong attractive interactions. Finally, the phase evolution allows for the measurement of an energetic equilibration timescale, describing the initial approach of the phase velocity to the polaron energy. Theoretically, this is shown to lie within the regime of universal dynamics revealing a fast initial evolution towards the formation of polarons. Our results give a comprehensive picture of the many-body physics governing the Bose polaron and thus validates the quasiparticle framework for further studies.
\end{abstract}

\maketitle

\section{Introduction}
The concept of quasiparticles is widely used in many areas of physics. It simplifies otherwise complex scenarios where interactions in a system can be described as emerging properties of quasiparticles instead. A canonical example is the polaron pioneered by Landau and Pekar~\cite{LandauPekar} which describes electrons coupled to lattice vibrations in crystals. However, these polarons have been difficult to explore systematically in condensed matter settings due to challenges such as high densities, fast evolution times, and disorder. In the past few years such quasiparticles have also been realized experimentally using quantum gases. These systems can serve as powerful platforms for simulating otherwise inaccessible regimes with high precision and have continuously advanced our understanding of interacting quantum systems~\cite{Bloch2012}. 

Initial studies focused on the properties of impurities in Fermi gases using spectroscopic methods~\cite{Schirotzek2009,Kohstall2012,Koschorreck2012,Scazza2017,Yan2019FermiPolaron,Darkwah2019,Fritsche2021} to measure the energy of the emergent Fermi polaron and its quasiparticle residue. Subsequently, investigations of the dynamical evolution of the Fermi polaron were conducted~\cite{cetina2015,cetina2016}, which provided evidence of a quantum beat between two polaron branches. However, the properties of mobile impurities in Bose-Einstein condensates (BECs) and the resulting Bose polaron are in some sense closer to condensed matter settings, since the bosonic nature of the bath mimics that of the crystal lattice surrounding electrons. The Bose polaron was first observed spectroscopically~\cite{jorgensen2016,hu2016,ardila2019,Yan2020} where the attractive and repulsive polaron branches were identified, providing a benchmark for the theoretical understanding of the polaron. Recently, the first experimental investigation of the dynamical properties of the Bose polaron was reported~\cite{Skou2021NatPhys}. Here, different regimes of dynamical behavior were observed from initial two-body universal dynamics through two-body weak coupling and finally dynamics governed by many-body correlations.

\begin{figure}[t!]
	\centering
	\includegraphics[width=1\columnwidth]{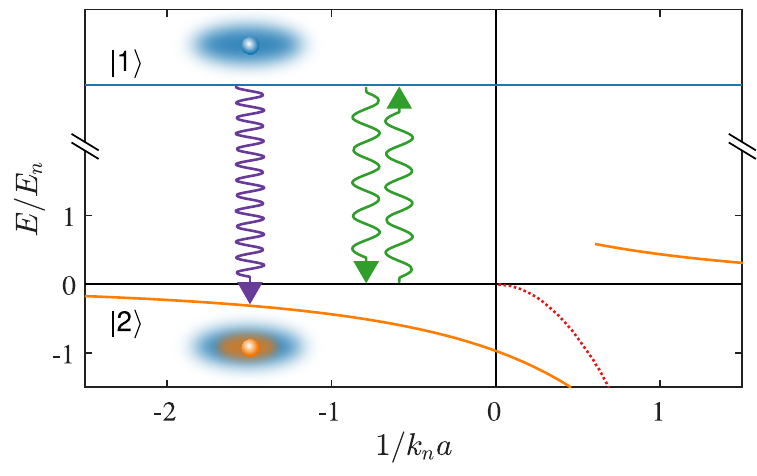}
	\caption{
		Illustration of the polaron energy spectrum and experimental methods. The energy of the medium $\ket{1}$ and impurity $\ket{2}$ states are shown in solid blue and orange lines respectively. The impurity state is shifted in energy away from its value in a vacuum (horizontal black line) due to interactions with the medium and features an attractive and a repulsive polaron branch. Spectroscopic methods use a single radio-frequency (rf) pulse (purple arrow) to investigate the spectral response of the impurity by transferring a few atoms into the impurity state whereby polarons are formed. Interferometric sequences use two short rf pulses (green arrows) to measure the dynamical evolution of a superposition of the medium state and the impurity state. The dotted red line shows the molecular state on the repulsive side of the Feshbach resonance. 
	}
	\label{IntroFig}
\end{figure}

In this paper, we present a comparison of interferometric and spectroscopic observations of the Bose polaron. Both methods use the same two quantum states as medium and impurity, and their respective energies are illustrated in Fig.~\ref{IntroFig}. The spectroscopic approach investigates the spectral response of the polaron through a three-body loss signal, and the interferometric method employs a Ramsey-like sequence to extract the dynamics of an impurity state as it interacts with the surrounding medium. Comparing the experimental pulse lengths with the underlying regimes of the impurity dynamics and the polaron lifetime reveals suitable regimes for interferometric and spectroscopic approaches. Finally, this work extends previous interferometric investigations~\cite{Skou2021NatPhys} to the regime of strong repulsive interaction and introduces the phase velocity as a tool to access the polaron energy in comparison with spectroscopic results ~\cite{jorgensen2016,ardila2019} in the system under investigation.

The paper is structured as follows. First, experimental details of the spectroscopic and interferometric observations and loss measurements of the polaron are discussed in Sec.~\ref{sec:Exp}, and their characteristic timescales are compared to the underlying dynamical regimes~\cite{Skou2021NatPhys}. Then, an analysis of the coherence amplitude based on the interferometric measurements is presented in Sec.~\ref{sec:NearUnitaryInt}. This includes the observation of the impurity dynamics at large repulsive interaction strengths. Finally, Sec.~\ref{sec:PhaseEvol} provides an analysis of the phase evolution using a new fitting procedure. This allows us to extract an interferometrically measured impurity energy from weak to strong attractive interactions, and to compare it with previous spectroscopic results. Furthermore, it enables us to analyze the initial evolution towards the Bose polaron state. 

\section{Experimental details}
\label{sec:Exp}

The energy spectrum of the Bose polaron has previously been investigated spectroscopically~\cite{jorgensen2016,hu2016,Yan2020}, and interferometric observations have measured its dynamical evolution from a bare impurity towards the formation of the polaron~\cite{Skou2021NatPhys}. In the following, we review the experimental details of these approaches \cite{jorgensen2016,Skou2021NatPhys}. By comparing to the underlying regimes of the impurity dynamics, we elucidate how these techniques can be used to probe different aspects of polarons.

Both the spectroscopic and interferometric experiments were performed in the same system using $^{39}$K BECs. These were produced in an optical dipole trap~\cite{wacker2015} in the $\ket{F=1, m_F = -1} \equiv \ket{1}$ hyperfine state. A second hyperfine state $\ket{F=1, m_F = 0} \equiv \ket{2}$ served as the impurity state, as indicated in Fig.~\ref{IntroFig}. The interaction between the two states is characterized by the dimensionless parameter $1/k_n a$, where the wave number $k_n = (6\pi^2 n_\text{B})^{1/3}$ is set by the average condensate density $n_\text{B}$ and $a$ is the scattering length between the two states. Importantly, $a$ can be controlled by magnetic fields through a Feshbach resonance located at $\SI{113.8}{G}$~\cite{lysebo2010,tanzi2018}, giving access to large interaction strengths. Moreover, the scattering length between medium atoms is constant $a_\text{B}\approx 9 a_0$ for the applied magnetic fields, where $a_0$ is the Bohr radius. Finally, the impurity-impurity scattering length is $\approx-15a_0$ and is not relevant due to low impurity concentrations.

\subsection{Spectroscopic measurements} 
\label{subsec:spectroscopy}
In a radio-frequency (rf) spectroscopic experiment, two states are coupled with a rf field. For the system at hand, these are the impurity and medium state with an approximate atomic transition frequency of $\SI{76}{MHz}$. Interactions between the two states lead to an energy shift of the resonance frequency, which is measured by varying the frequency of the rf pulse. The experiment is then repeated for many values of $1/k_na$ thus mapping out the complete spectral response of the impurity. Spectroscopic measurements generally require a long probe pulse of low power to resolve the spectrum, whose relevant energy scale is set by $E_n = \hslash^2k_n^2/2m$, where $m$ is the mass of $^{39}$K. A square pulse of $\SI{100}{\micro s}$ duration was, therefore, used in Ref.~\cite{jorgensen2016}, achieving an experimental resolution of $0.15 E_n$. 

If the frequency of the rf pulse matches the transition frequency from the medium state to the impurity state, atoms are transferred into this state and polarons are formed in the medium, as indicated in Fig.~\ref{IntroFig}. The transition frequency is generally broadened due to the continuum of excited states and the inhomogeneous density of the medium. Following the formation of polarons, they are quickly lost through three-body recombination with two medium atoms. It is therefore not possible to obtain a direct measurement of the polarons. However, the loss mechanism can be utilized as the experimental signal. By varying the frequency of the rf pulse while measuring the number of medium atoms, the spectral response of the polaron can be extracted. This procedure is an example of injection spectroscopy and probes both the ground state of the polaron and the continuum of excited states. For the specific measurements in Ref.~\cite{jorgensen2016}, the pulse power was chosen to transfer $10\%$ into the impurity state. 

\subsection{Interferometric measurements} 
\label{subsec:Interferometry}
To investigate the formation of the Bose polaron, a different technique is required to resolve the dynamical evolution. This can be achieved with an interferometric sequence, which produces a coherent superposition with a first rf pulse and probes the system with a second rf pulse following a variable evolution time, corresponding to a Ramsey sequence~\cite{Scelle2013,cetina2015,cetina2016,Rentrop2016}. However, instead of the usual $\pi/2$ pulse leading to equal populations, a small initial rotation is employed. This allows the $\ket{1}$- and $\ket{2}$-state to be assigned as medium and impurity state, respectively. In the experimental system outlined above, this procedure can be realized with very short pulses of $\SI{0.5}{\micro s}$ duration resonant with the bare atomic transition, as shown by the green arrow in Fig.~\ref{IntroFig}. This corresponds to $\pi/7$ pulses creating a superposition of the $\ket{1}$- and the $\ket{2}$-state with a $\sim 5\%$ admixture of the latter. After a variable hold time, $t$, a second pulse probes the impurity dynamics towards the formation of the polaron. This pulse is applied with a variable phase between $0$ and $2\pi$. Depending on this phase, either additional atoms are transferred to the impurity state or impurity atoms are transferred back to the medium state, resulting in a sinusoidal Ramsey signal. Following the interferometric sequence, absorption imaging is used to extract the spin population and by fitting to the measured Ramsey signal, we extract the coherence function $C(t) = |C(t)|e^{i\varphi_C(t)}$. The latter can be obtained from the impurity Green's function $G(t) = -iC(t) = -i\bra{\psi_{\rm BEC}}\hat{c}(t)\hat{c}^\dagger(0)\ket{\psi_{\rm BEC}}$, where $\ket{\psi_{\rm BEC}}$ describes the state of the BEC before the first rf pulse and $\hat{c}^\dagger$ is the operator that creates an impurity in the condensate. Interactions cause the system to decohere characterized by $|C(t)|$ and to evolve with a certain phase $\varphi_\text{C}(t)$. 

\begin{figure}[t!]
	\centering
	\includegraphics[width=1\columnwidth]{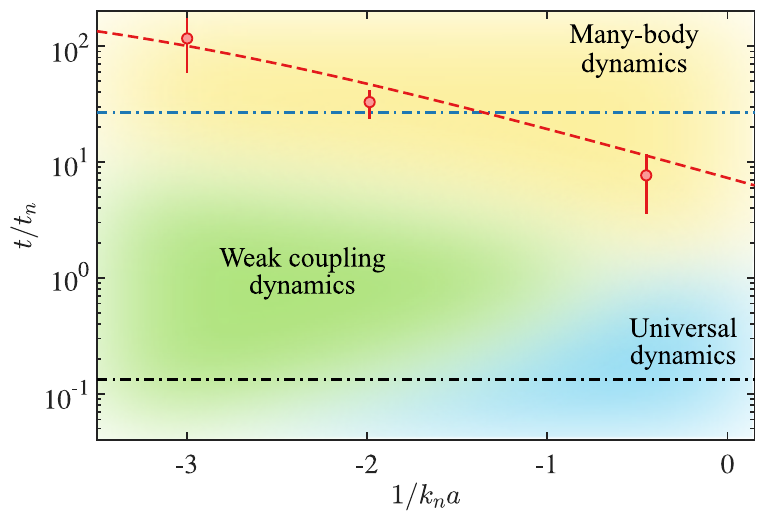}
	\caption{Experimental pulse lengths, polaron lifetime and dynamical regimes in units of $t_n$. The black and blue dashed-dotted lines show the interferometric and spectroscopic pulse lengths respectively. The red circles are impurity lifetime measurements with a phenomenological fit shown as the dashed red line~\cite{Skou2021NatPhys}. These are compared to the underlying dynamical regimes -- universal, weak coupling and many-body -- shown as the colored background. }
	\label{Pulselengths_vs_dynamical_regimes}
\end{figure}

In Ref.~\cite{Skou2021NatPhys}, three regimes of dynamical behavior were identified, as indicated in Fig.~\ref{Pulselengths_vs_dynamical_regimes}. At short times, only high energy unitarity limited scattering contribute significantly to decoherence, which gives rise to universal dynamics depending on the condensate density only, shown as the blue region in Fig.~\ref{Pulselengths_vs_dynamical_regimes}. At weak coupling, there is a crossover around $t_a = ma^2/\hslash$ to another regime, where the important decoherence processes are governed by scattering events with a cross-section proportional to the impurity-medium scattering length squared, corresponding to the green region in Fig.~\ref{Pulselengths_vs_dynamical_regimes}. Finally, the system transitions into a regime of many-body dynamics designated by the yellow region, leading to the formation of the polaron. At weak coupling, this transition scales inversely with the mean-field energy $E_\text{mf} = 4\pi \hslash^2 n_\text{B} a / m$ and decreases with increasing interaction strength. However, it remains finite at strong interactions with a value of $\approx 1.4 t_n$, where $t_n = \hslash /E_n \approx \SI{3.7}{\micro s}$, and describes a direct crossover from the universal short-time dynamics to the many-body regime.

Crucially, we note that the interferometric $\pi / 7$ pulses are typically shorter than the timescale of these dynamical regimes, showing that the interferometric measurements can resolve the dynamics of the impurity even at the Feshbach resonance, where the dynamics occurs at the unitarity-limited timescale $t_n$.

\subsection{Impurity lifetime} \label{subsec:lifetime}
The impurity lifetime sets an important limit for both spectroscopic and interferometric measurements, and ultimately leads to the decay of the polaron. Most importantly, the lifetime of the impurity should be longer than the duration of the dynamical evolution towards the polaron state such that it constitutes a well-defined quasiparticle. We therefore investigate this lifetime  experimentally following Ref.~\cite{Skou2021NatPhys}. At a chosen interaction strength, a BEC is prepared in the $\ket{1}$ state and an initial rf pulse of $\SI{0.8}{\micro s}$ duration is used to create a superposition of the $\ket{1}$ and $\ket{2}$ states corresponding to a $10\%$ population in the latter. In a following evolution time $t$, some of these impurities are lost predominantly due to three-body collisions, where two medium atoms are lost for each impurity atom. Subsequently, a second rf pulse of $\sim \SI{9}{\micro s}$ duration transfers the remaining impurity atoms to a third hyperfine state $\ket{F=1, m_F =+1} \equiv \ket{3}$. The impurities in the $\ket{3}$ state can perform spin-flip collisions with the medium atoms in the $\ket{1}$ state, where both atoms obtain sufficient kinetic energy to leave the trap. This fast two-body mechanism removes a single medium atom for each impurity atom practically instantaneously. Since the two processes lead to a different number of lost medium atoms, this sequence allows us to obtain the lifetime of impurities in state $\ket{2}$ by measuring the number of medium atoms in state $\ket{1}$, as a function of the evolution time $t$. This lifetime at different interaction strengths is shown as red points in Fig.~\ref{Pulselengths_vs_dynamical_regimes}. Following Ref.~\cite{Skou2021NatPhys}, we also perform a phenomenological fit of the lifetime in the form $t_{\rm l}^{-1} = \beta_1 + \beta_2 \exp(\beta_3 / k_n a)$ with fitting parameters $\beta_i$, across all attractive interaction strengths, $k_n a < 0$. This is shown as the red dashed line in Fig.~\ref{Pulselengths_vs_dynamical_regimes}. This analysis confirms the desired separation of timescales, since the lifetime is longer than the transition time to the many-body regime. Such a separation of timescales leads to a well-defined polaron and enables its investigation. 

In spectroscopic studies of the polaron, the ideal scenario is a pulse length longer than the transition time to the many-body regime, such that the polaron has time to be formed, but shorter than the lifetime. However, a single pulse length may not necessarily fulfill this requirement at all interaction strengths. In Fig.~\ref{Pulselengths_vs_dynamical_regimes}, the compromise for the pulse length chosen for the spectroscopic experiments in Ref.~\cite{jorgensen2016} is shown. For strong interactions, it is limited by the impurity lifetime and in future experiments an adaptive pulse length could be employed. Specifically, Fig.~\ref{Pulselengths_vs_dynamical_regimes} shows that longer pulses are optimal for investigations at weak interaction strengths, whereas shorter pulses are well suited at strong interactions.


\section{Near-unitary impurity dynamics}
\label{sec:NearUnitaryInt}
In this section, we analyze the interferometrically measured coherence amplitude $|C(t)|$ close to and on both sides of the unitary limit, $1 / k_n|a| = 0$. Figure~\ref{Pulselengths_vs_dynamical_regimes} shows that at these large interaction strengths, we expect the dynamics to transition directly from an initial universal behavior depending only on the condensate density into a regime of many-body correlations~\cite{Skou2021NatPhys}, in which the polaron forms. 

\begin{figure}[t!]
\centering
		\includegraphics[width=\columnwidth]{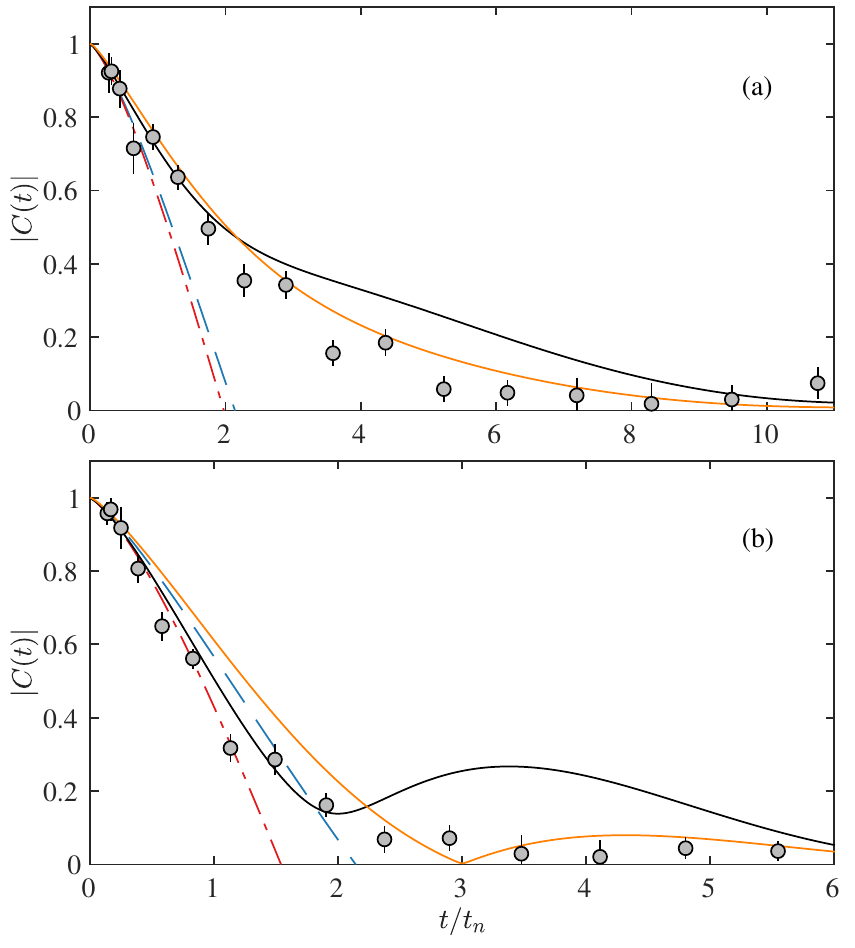}
		\caption[d]{ Impurity dynamics at near-unitary interaction strengths. The measured coherence amplitudes at $1/k_na = -0.23$ (a) and $1/k_n a = +0.29$ (b) are shown as gray circles. At short times, the amplitude is governed by the unitarity-limited universal behavior of Eq.~\eqref{eq:Amp_2body} up to order $t^{3/2}$ shown as dashed blue lines, i.e. setting $k_2 = 0$. The second order contribution is included in the dash-dotted red lines and contains the next-order correction with $k_2 = 0.03$ for $1/k_na = -0.23$ and $k_2 = 0.165$ for $1/k_n a = +0.29$. The diagrammatic prediction, which includes both polaron branches, is shown as solid lines. The black lines show the result of the non-selfconsistent approach, while the orange lines correspond to the modified approach, where the unphysical gap between the attractive polaron branch and the many-body continuum is removed.}
		\label{SingleFigInt}
\end{figure}

The observed coherence amplitudes are shown in Fig.~\ref{SingleFigInt} and generally exhibit a fast initial decrease followed by a slower decay for longer evolution times~\cite{NewData}. Intriguingly, the dynamics observed for a large repulsive interaction with $1/k_na=0.29$ shown in panel (b) clearly evolves \emph{faster} than the dynamics for a large and attractive interaction $1/k_na=-0.23$ shown in panel (a). This key finding reveals the difference between impurity dynamics at large attractive and repulsive interaction strengths, otherwise similar in magnitude, and is investigated further in the following.

\subsection{Initial dynamics}
We begin by analyzing the interferometric observations for short evolution times, $t\lesssim t_n$. Theoretically, the coherence dynamics can be obtained from the Fourier transform of the spectral function. For times $t \ll ma^2 / \hslash$, the resulting short-time dynamics can be expanded in orders of $t / t_n$ giving
\begin{align}
|C(t)| = 1 - k_{3/2} \left(\frac{t}{t_n}\right)^{3/2} - k_2 \left(\frac{t}{t_n}\right)^{2}.
\label{eq:Amp_2body}
\end{align}
The dominant $t^{3/2}$ behavior originates from the tail of the spectral function for high frequencies~\cite{braaten2010}, and is a direct consequence of the fact that short time decoherence is governed by high energy unitarity limited scattering. This leads to the universal value of $k_{3/2} = 16 / 9\pi^{3/2}$. The universal prediction up to order $t^{3/2}$ has previously been observed to capture impurity dynamics directly at unitarity~\cite{Skou2021NatPhys}, and indeed this prediction compares very well to the initial dynamics for $1 / k_na = -0.23$ shown in Fig.~\ref{SingleFigInt}(a). However, the dynamics at repulsive interactions evolves faster than this density-limited prediction~\cite{Decoherence}.

To understand this difference in initial coherence dynamics for positive and negative interaction strengths, we extend the analysis to include the next-order correction. The coefficient $k_2$ in front of the second-order term is obtained by taking the presence of the attractive polaron branch into account (see Appendix ~\ref{app:second_order_coeff} for details). This results in the approximate value of $k_2 = Z_\text{p} (E_\text{p} / E_n)^2 / 2 - 4 / (3\pi k_n |a|)$. The attractive polaron peak contributes to $k_2$ with the term $Z_\text{p} (E_\text{p} / E_n)^2 / 2$ depending on its quasiparticle residue $Z_\text{p}$ and energy $E_\text{p}$, which tends to \emph{speed up} the dynamics. The quasiparticle properties are obtained from a non-perturbative diagrammatic approach based on the ladder approximation~\cite{Rath2013} at zero temperature. On the contrary, the term $- 4 / (3\pi k_n |a|)$ stemming from the high-frequency tail of the spectral function tends to \emph{slow} the dynamics down. The combination of these effects yields good agreement with the experimental observations, as explicitly shown in Fig.~\ref{SingleFigInt}. Thus, the faster evolving coherence amplitude found for $1 / k_na = +0.29$ is due to a larger magnitude of the polaron energy $|E_\text{p}|$ on the attractive branch evident from Fig.~\ref{IntroFig}, as it smoothly connects to the two-body bound state at $-\hslash^2 / ma^2$. 
	
\subsection{Many-body regime}
For large interaction strengths $|1/k_na|\lesssim (2/3\pi)^{1/3}\approx 0.6$~\cite{Skou2021NatPhys}, the impurity dynamics transitions directly from universal two-body behavior into the regime of many-body correlations, as illustrated in Fig.~\ref{Pulselengths_vs_dynamical_regimes}. At these strong interactions, there exists no exact solution for the many-body dynamics, and it remains subject of active theoretical investigations including diagrammatic~\cite{Rath2013,Casteels2014,Christensen2015,Camacho-Guardian2018b}, variational ~\cite{Tempere2009,Levinsen2015,Shchadilova2016,Guenther2018,Liu2019,ardila2021,Drescher2021,Massignan2021,christianen2021} and quantum Monte Carlo~\cite{Ardila2015,Ardila2016,Chen2018} approaches. Here, we employ a diagrammatic approach based on the so-called ladder approximation~\cite{Rath2013}, which accounts exactly for the Feshbach physics to calculate the spectral function of the impurity. Similar to the short-time predictions, the impurity coherence is then obtained by Fourier transforming this spectral function for a given interaction strength~\cite{Skou2021NatPhys}. The results of this calculation are plotted in Fig.~\ref{SingleFigInt}~\cite{Decoherence} (black lines). This approach, however, leads to a gap between the attractive polaron branch and the continuum of states lying above~\cite{Rath2013}. This is unphysical, since one can generate states with arbitrarily small excitation energy consisting of a moving polaron and a Bogoliubov mode with a total momentum of zero. Therefore, we modify the diagrammatic prediction by closing the gap as outlined in Appendix ~\ref{app.ladder_approximation}. In the case of strong attractive interactions, shown in Fig.~\ref{SingleFigInt}(a), the observations are captured well by this modified diagrammatic description (orange lines). This remarkable agreement of the experimental observations and the modified ladder approximation shows that the main characteristics of the many-body problem can be understood on these grounds. 
 
On the other side of the resonance, the repulsive interactions lead to two polaron branches as shown in Fig.~\ref{IntroFig}. The attractive branch crosses the resonance approaching the molecular state and at higher energies a damped repulsive state emerges and becomes well-defined for $1/k_na\approx+0.5$. Both branches are captured by the diagrammatic description, and the calculated coherence consequently evolves faster for repulsive interactions, as shown in Fig.~\ref{SingleFigInt}(b), in good agreement with the observations. For intermediate times, $t_n < t < 2t_n$, the modified diagrammatic approach predicts slightly slower dynamics compared to the experimental data and the approximate short-time behavior in Eq.~\eqref{eq:Amp_2body}, as well as the non-selfconsistent ladder approximation. This indicates that the resulting discrepancy is mainly due to the method of removing the gap between the attractive polaron branch and the continuum (Appendix \ref{app.ladder_approximation}).
		
Another special feature at repulsive interactions is a quantum beat between the two polaron branches~\cite{cetina2016}. For the coherence amplitude, this revival can clearly be seen in the theoretical prediction, despite the fact that the quantum beat between the attractive polaron and the emergent repulsive branch is smoothened by the trap average, which is treated at the level of the local density approximation [see Appendix \ref{app.ladder_approximation}]. In the experimental realization, such a revival is suppressed by the decay of the repulsive polaron, three-body losses, and decoherence processes induced by the harmonic trapping potential and magnetic shot-to-shot noise. However, the data tentatively displays a minimum followed by a small revival around $\approx 5t_n$, which may indicate such a quantum beat, between the attractive polaron and the emerging repulsive state. 
		
These observations constitute the first measurement of impurity dynamics at repulsive interaction strengths in a BEC and open a door to unveiling the intriguing interplay between the emergent polaron states responsible for the effective dynamics.

\section{Phase evolution}
\label{sec:PhaseEvol}
The phase evolution of a quantum state is generally a rich source for obtaining information of a given system, and has previously been used to obtain the two- and three-body contact in a unitary Bose gas~\cite{fletcher2017}. In the context of impurities in a bosonic medium, it was previously shown~\cite{Skou2021NatPhys} that the short-time dynamics of the phase has a power-law behavior with a characteristic exponent corresponding to the regime of universal or weak coupling dynamics. In the following, we extend the analysis of the phase evolution, and show that it can be used to extract the polaron energy and thus characterize the initial evolution towards the polaron state.

\subsection{Polaron energy}
In this section, we extract the energy of the polaron from the measured phase evolution. This permits a comparison between the interferometrically inferred polaron energy and previous spectroscopic results.

Like the coherence amplitude, the coherence phase generally evolves according to the different dynamical regimes~\cite{Skou2021NatPhys,Skou2021MDPI} shown in Fig.~\ref{Pulselengths_vs_dynamical_regimes}. However, at long evolution times, when the polaron has formed, the phase is expected to evolve according to the polaron energy. In this regime the phase evolution is therefore expected to be linear $\varphi_C(t) \to -E_\text{p}\cdot t/\hslash$. This is similar to the phase of a two-level system in a Ramsey sequence, which evolves with fixed speed for a detuning $\Delta$ between the frequency of the probing pulse and the transition frequency. For long evolution times, the average phase velocity, $v_{\rm ph} = \hslash \varphi_C(t) / t$, thus approaches a constant corresponding to the polaron energy $\Delta = -E_\text{p}$, and consequently reveals $E_\text{p}$. This allows us to extract the polaron energy from the interferometric measurements, which can then be compared to previous spectroscopic studies~\cite{jorgensen2016}.

\begin{figure}[t]
	\centering
	\includegraphics[width=1\columnwidth]{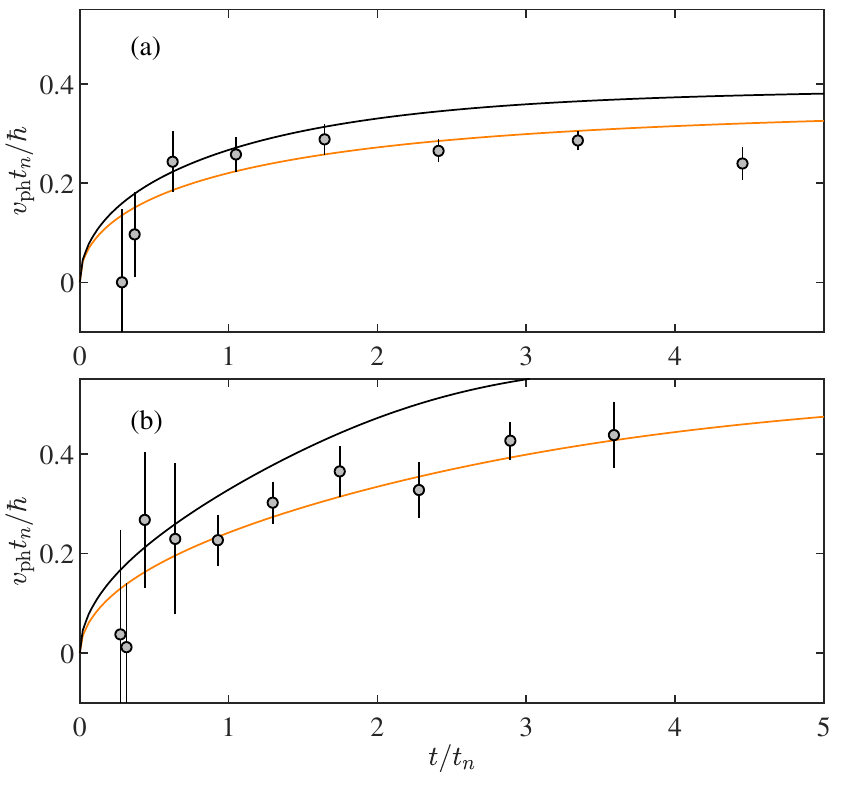}
	\caption{Phase velocity $v_{\rm ph} = \hslash \varphi_C(t) / t$. The phase velocity is shown for (a) $1/k_n a = -0.67$ and (b) $1/k_n a = -0.23$ with data indicated as circles and the modified and non-selfconsistent diagrammatic prediction shown in solid orange and black lines respectively.}
	\label{PhaseVel}
\end{figure}

Examples of measured phase velocities are shown in Fig.~\ref{PhaseVel}~\cite{NewData}. At intermediate interaction strengths, we observe the expected saturation of the phase velocity, as shown in Fig.~\ref{PhaseVel}(a). At stronger interactions, however, the evolution is slower and the phase velocity has not yet saturated completely at the latest observable times before the coherence is lost, as shown in Fig.~\ref{PhaseVel}(b). In general, we observe remarkable agreement of the modified diagrammatic predictions, shown in solid orange lines, with experimental observations, whereas the non-selfconsistent diagrammatic approach only qualitatively agrees.

\begin{figure}[t]
	\centering
	\includegraphics[width=1\columnwidth]{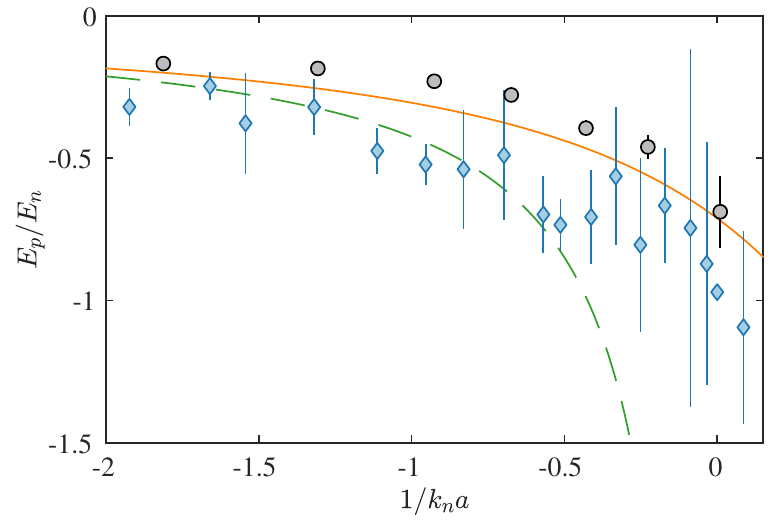}
	\caption{Polaron energy at attractive interaction strengths. The fitted energies from the interferometric measurements are shown as gray circles, while the spectroscopically measured polaron energies~\cite{ardila2019} are shown as blue diamonds. Our diagrammatic prediction and the mean field energy are shown as a solid orange line and a dashed green line respectively. The error bars indicate the standard error of the fit to Eq. \eqref{eq.phase_velocity} and the root-mean-square spectral width for the interferometric and spectroscopic data respectively.}
	\label{ImpEnergy}
\end{figure}

To extract the polaron energy, we perform a phenomenological exponential fit
\begin{equation}
v_{\rm ph}(t) = -E_\text{p} \cdot \left[ 1 - \exp\left(-t/\tau \right)  \right],
\label{eq.phase_velocity}
\end{equation}
to the phase velocity. Here, the polaron energy $E_\text{p}$ and the timescale $\tau$ are independent fit parameters. Figure~\ref{ImpEnergy} shows the extracted average polaron energies together with previous spectroscopic observations of the polaron energy at maximum density~\cite{ardila2019}, a diagrammatic prediction and the mean-field energy $E_\text{mf} = 4\pi\hslash^2n_\text{B} a /m$. The interferometrically measured energies decrease with increasing interaction strength and the most bound polaron at unitarity has an energy of $\approx -0.7 E_n$. Generally, the interferometric data agrees well with the diagrammatic prediction, especially at unitarity. The spectroscopic result is in qualitative agreement with this prediction and clearly favors the polaron energy for strong interactions. This is in agreement with the results reported in Refs.~\cite{jorgensen2016,ardila2019}, where it was found that including correlations beyond the ladder approximation led to an improved agreement with the spectroscopic data. 

Note that Fig.~\ref{ImpEnergy} displays the spectroscopically measured polaron energy at \emph{maximum} density~\cite{ardila2019} and the \emph{average} polaron energies from the interferometric measurements. To account for this difference, the value $k_n = (6\pi^2 n_\text{B})^{1/3}$ is calculated using the maximal and average density respectively. This, however, assumes that the polaron energy in units of $E_n$ is only a function of the interaction strength $1 / k_n a$. While additional effects, e.g. due to the presence of Efimov trimers are known to influence the polaron energy~\cite{Levinsen2015,Yoshida2018}, we assume these effects to be small.

In general, we note that the interferometrically inferred energies are consistently higher than both the spectroscopic measurements and the theoretical predictions. This may be a result of the fitting procedure, which is likely to underestimate the magnitude of the polaron energy, $|E_\text{p}|$, or stem from unaccounted loss processes, which would slow down the phase evolution for long times. However, the overall good agreement provides evidence that the interferometric and spectroscopic methods inherently create and probe the same physics, though they differ in experimental approach with pulse lengths separated by orders of magnitude. 

\subsection{Energetic equilibration timescale}
Theoretically, the formation timescale can be defined as the time when the impurity coherence amplitude is equilibrated at the quasiparticle residue~\cite{Shchadilova2016,nielsen2019}. In experimental realizations, this equilibration is often masked by decoherence mechanisms such as dephasing from inhomogeneous density distributions, finite impurity lifetime and magnetic field fluctuations, prohibiting a direct observation of this timescale. It is, however, possible to determine the transition between the different regimes of the impurity dynamics [Fig.~\ref{Pulselengths_vs_dynamical_regimes}] both theoretically and experimentally~\cite{Skou2021NatPhys}. Our present analysis additionally provides an equilibration timescale for the phase evolution, which can add to the understanding of these dynamical regimes.

\begin{figure}[t!]
	\centering
	\includegraphics[width=1\columnwidth]{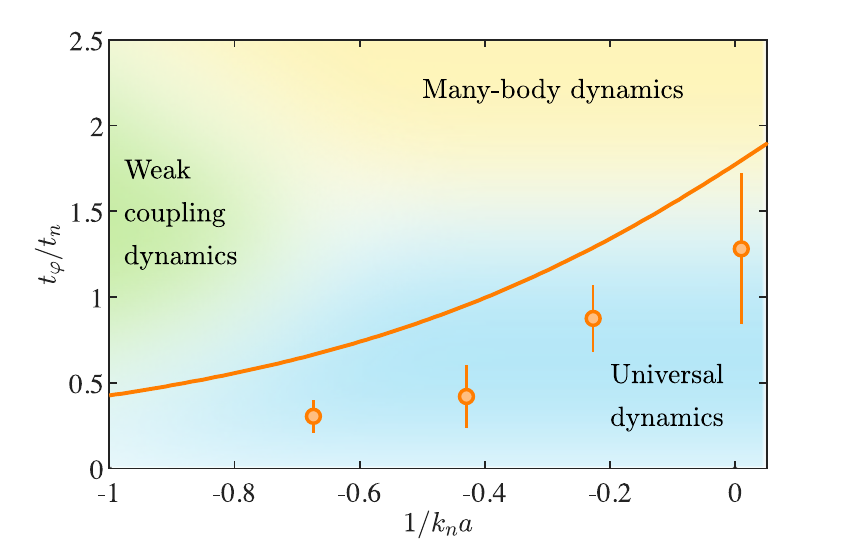}
	\caption{Equilibration timescale $t_\varphi$ and dynamical regimes. The energetic equilibration timescale of the phase evolution from the diagrammatic prediction is shown in orange lines, whereas the experimentally extracted timescale is shown in points. These are defined as the halfway time of the phase velocity, i.e. the time at which the phase velocity is half the way to the polaron energy. Finally, these are compared to the dynamical regimes (colored background) also shown in Fig.~\ref{Pulselengths_vs_dynamical_regimes}.}
	\label{EqTime}
\end{figure}

Figure~\ref{EqTime} shows the energetic equilibration timescale $t_{\varphi}$. We define this as the halfway time of the phase velocity, i.e. the time at which the phase velocity is equal to half the polaron energy. Theoretically, this is computed within the diagrammatic approach. Experimentally, it is extracted from the exponential fits according to  Eq.~\eqref{eq.phase_velocity} as $t_\varphi = \tau \cdot \ln2$. Both of these are then compared to the dynamical regimes shown in the background of Fig.~\ref{EqTime}. 

For weak interactions, the theoretically extracted timescale closely follows the transition from the universal to weak coupling regime, $t_\varphi \sim t_a = ma^2 / \hslash$. This is expected, since the mean field energy develops for $t \gtrsim t_a$~\cite{Skou2021NatPhys}. At intermediate to strong interaction strengths, the equilibration timescale is less than or similar to the transition to the many-body regime. This indicates that across all interaction strengths, the phase velocity initially evolves quickly towards the polaron energy. In the regime of many-body dynamics, the final approach to the polaron energy takes place, which is a much slower process than the initial dynamics in line with what has been observed in the perturbative limit~\cite{nielsen2019}. 

Thus our analysis of the phase evolution of the impurity coherence dynamics provides further understanding of the evolution towards the Bose polaron. Although Fig.~\ref{EqTime} shows a clear qualitative agreement between the theoretical prediction and the timescales extracted from the fitting function in Eq.~\eqref{eq.phase_velocity}, a quantitative difference between the two remains. Since our diagrammatic approach actually describes the experimentally inferred phase velocities very well as shown by Fig.~\ref{PhaseVel}, this discrepancy is mainly an artifact of the phenomenological fitting function [Eq.~\eqref{eq.phase_velocity}] which underestimates the magnitude of the polaron energy, $|E_\text{p}|$. Nonetheless, our simple approach lends further intuition into the timescales involved in the equilibration of the Bose polaron.

\section{Conclusion}
To summarize, we have provided a detailed investigation of interferometric and spectroscopic measurements of the Bose polaron. The results include an analysis of the timescales of spectroscopic and interferometric sequences with respect to the lifetime, an observation of impurity dynamics at repulsive interaction strengths, and an extraction of the polaron energy from interferometric observations in agreement with previous results.

The comparison of timescales showed a separation between the transition to the many-body regime and the polaron lifetime at all attractive interaction strengths, indicating a well-defined quasiparticle. We also compared the interferometric and spectroscopic pulse lengths to the underlying dynamical regimes of the impurity dynamics and the polaron lifetime. This highlights the capability of the interferometric approach to clearly resolve polaron dynamics and suggests new routes for spectroscopic measurements.

Secondly, an interferometric measurement of the coherence amplitude at strong interactions revealed faster impurity dynamics and polaron formation at large repulsive interaction strengths than at corresponding attractive interactions. For short times, the observations were in excellent agreement with a short-time prediction including both the continuum and the attractive polaron branch. For longer times, a qualitative agreement was obtained with a many-body theoretical prediction which included both branches. This prompts the necessity for further investigations of impurity dynamics at repulsive interaction strengths.

Thirdly, we obtained the polaron energy from interferometric measurements by fitting to the observed phase velocities. This energy was compared with previous spectroscopic results from weak to strong attractive interactions obtaining qualitative agreement. Finally, the phase evolution also allowed us to define an energetic equilibration timescale, describing the initial approach of the phase velocity to the polaron energy. Theoretically, this was shown to closely follow the transition from the universal to the weak coupling regime at weak interactions, and approaching $\sim t_n$ in the unitary limit. A similar analysis of the phase velocity at various repulsive interaction strengths based on further interferometric investigations will be the topic of future work.

Based on this thorough discussion of the two experimental approaches employed so far, the stage is set for further investigation of the Bose polaron. While the size of the smallest Efimov trimer is much larger than the interparticle spacing, it may nonetheless influence the polaron energy~\cite{Levinsen2015,Yoshida2018}. Expanding investigations of the polaron energy at unitarity, by varying the density, may enable studying such a universal dependence on the Efimov three-body parameter. Another interesting aspect is the dynamics at higher impurity fractions, which may show mediated polaron interactions~\cite{Camacho-Guardian2018b}. Such effective interactions are predicted to enable deeply bound states of bosonic bipolarons~\cite{Camacho-Guardian2018}, which have so far been elusive to experimental observation. However, employing ejection spectroscopy~\cite{Yan2020} may enable measuring these exotic states.

\section{Acknowledgements}
This work was supported by the Danish National Research Foundation through the Center of Excellence (Grant Agreement No.: DNRF156), and the Independent Research Fund Denmark--Natural Sciences via Grant No. DFF -8021-00233B.
\appendix

\section{Derivation of second order coefficient} 
\label{app:second_order_coeff}
In this appendix, we derive the second order coefficient in Eq. \eqref{eq:Amp_2body}. The coherence $C(t)$ is in general the Fourier transform of the spectral function, which may be rewritten as
\begin{align}
C(t) = &\int_{-\infty}^{\infty} \frac{{\rm d}\omega}{2\pi} e^{-i\omega t} A(\omega) \nonumber \\
= &\int_{-\infty}^{\infty} \frac{{\rm d}\omega}{2\pi} (1 - i\omega t) A(\omega) + \nonumber \\ 
&\int_{-\infty}^{\infty} \frac{{\rm d}\omega}{2\pi} \left[e^{-i\omega t} - (1 - i\omega t)\right] A(\omega) \nonumber \\
\simeq\;  &1 - it \cdot \frac{n_\text{B} \mathcal{T}_\text{B}}{\hslash}\left(1 - \frac{a_\text{B}}{a}\right) + \nonumber \\ 
&\int_{-\infty}^{\infty} \frac{{\rm d}\omega}{2\pi} \left[e^{-i\omega t} - (1 - i\omega t)\right] A(\omega) 
\label{eq.coherence_from_spectral_function}
\end{align}
We use the sum rules \cite{braaten2010} $\int_{-\infty}^{\infty} {\rm d}\omega \, A(\omega) / 2\pi = 1$, $\int_{-\infty}^{\infty} {\rm d}\omega \, \omega A(\omega) / 2\pi = (a_\text{B}^{-1} \!-\! a^{-1})\hslash C_2 / (4\pi m N_\text{B}) \!\simeq\! 4\pi n_\text{B}a_\text{B} / m$ to compute the first term in Eq. \eqref{eq.coherence_from_spectral_function}. In the latter sum rule, we have used the two-body contact for the BEC, $C_2 \simeq N_\text{B} \cdot 16\pi^2 n_\text{B}a_\text{B}^2$, and the zero-energy $\mathcal{T}$-matrix for two mediumn atoms: $\mathcal{T}_\text{B} = 4\pi\hslash^2a_\text{B} / m$. To compute the remaining integral, we include the exact high-frequency tail \cite{braaten2010} ($t_a = ma^2 / \hslash$)
\begin{equation}
A_\text{tail}(\omega)=\lim_{\omega\to\infty} A(\omega) = \frac{1}{2\pi}\frac{C_2}{N_\text{B}}\sqrt{\frac{\hslash}{m}}\frac{\left(\frac{a}{a_\text{B}} - 1\right)^2}{(1 + \omega t_a)\omega^{3/2}},
\label{eq.high_frequency_tail} 
\end{equation}
as well as the quasiparticle peak around the polaron energy 
\begin{equation}
A_\text{p}(\omega) = A(\omega \sim E_\text{p}) = 2\pi Z_\text{p} \cdot \delta(\omega - E_\text{p} / \hslash). 
\end{equation}
Because of the factor $e^{-i\omega t} - (1 - i\omega t)$ in the integral in Eq. \eqref{eq.coherence_from_spectral_function}, this will yield the exact short-time dynamics up to, but excluding, order $t^{2}$. As we include the quasiparticle peak at low energies, we expect this to yield the dominant contribution at order $t^2$ as well. Explicitly, we may then write
\begin{align}
&\int_{-\infty}^{\infty} \frac{{\rm d}\omega}{2\pi} \left[e^{-i\omega t} - (1 - i\omega t)\right] A(\omega) \nonumber \\
=&\int_{-\infty}^{\infty} \frac{{\rm d}\omega}{2\pi} \left[e^{-i\omega t} - (1 - i\omega t)\right] \left[ A_\text{tail}(\omega) + A_\text{p}(\omega)\right] \nonumber \\
=& \; C_\text{tail}(t) + C_\text{p}(t). \nonumber
\end{align}
For the contribution from the quasiparticle peak, this yields
\begin{align}
C_\text{p}(t) &= \int_{-\infty}^{\infty} \frac{{\rm d}\omega}{2\pi} \left[e^{-i\omega t} - (1 - i\omega t)\right]  A_\text{p}(\omega) \nonumber \\
&= Z_\text{p} \left[e^{-i E_\text{p} t / \hslash} - (1 - i E_\text{p} t / \hslash)\right] \nonumber \\
&\simeq -\frac{Z_\text{p}}{2} \left(\frac{E_\text{p}}{E_n}\right)^2 \cdot \left(\frac{t}{t_n}\right)^2.
\label{eq.C_p}
\end{align}
The lower line is the second order expansion, in which we have used $t_n = \hslash / E_n$. The remaining contribution from the asymptotic tail of the spectral function yields
\begin{align}
\!\!C_\text{tail}(t) &= \frac{2\left(1 - \frac{a_\text{B}}{a}\right)^2}{3\pi} (k_n|a|)^3 \left[1 + i\frac{t}{t_a} - \frac{2e^{it/t_a}}{\sqrt{\pi}} \Gamma\left(\frac{3}{2}, i\frac{t}{t_a}\right) \right] \nonumber \\
 &\simeq 1 - (1 - i)\frac{16}{9\pi^{3/2}}\left(\frac{t}{t_n}\right)^{3/2} + \frac{4}{3\pi k_n|a|}\left(\frac{t}{t_n}\right)^2\!\!.\!\!
 \label{eq.C_asymp}
\end{align}
We expand the expression to second order in $t / t_a$ to get the lower line. Here, we also approximate $1 - a_\text{B} / a \simeq 1$, using that the scattering length between medium atoms, $a_\text{B}\simeq 9a_0$, in the present experimental setup is several orders of magnitude smaller than any of the studied impurity-medium scattering lengths, $a$. Inserting Eqs. \eqref{eq.C_p}-\eqref{eq.C_asymp} in \eqref{eq.coherence_from_spectral_function}, we finally obtain
\begin{align}
C(t) &\simeq 1 - it \cdot \frac{n_\text{B} \mathcal{T}_\text{B}}{\hslash}\left(1 - \frac{a_\text{B}}{a}\right)  + C_\text{tail}(t) + C_\text{p}(t) \nonumber \\
&\simeq 1 - (1 - i) k_{3/2}\left(\frac{t}{t_n}\right)^{3/2} - k_2 \left(\frac{t}{t_n}\right)^2,
\label{eq.C_short_times_result}
\end{align}
with the universal coefficient, $k_{3/2} = 16/9\pi^{3/2}$, and the approximate second order coefficient $k_2 = Z_\text{p} (E_\text{p} / E_n)^2 / 2 - 4/(3\pi k_n|a|)$. Again, the scattering length for two medium atoms is only $a_\text{B} \simeq 9a_0$ in the present experiments, and, therefore, the linear term would only be visible at exceedingly small timescales. Consequently, we drop this term in the lower line as well as in Eq. \eqref{eq:Amp_2body} of the main text. 

\section{Ladder approximation and gap closing} \label{app.ladder_approximation}
In this appendix, we briefly outline the many-body formalism based on the ladder approximation to the impurity-medium $\mathcal{T}$-matrix \cite{Rath2013}, which includes Feshbach physics via the scattering of one boson out of the condensate by the impurity. Throughout the paper, spectral functions are always evaluated at zero momentum. \\

In the experiment, the medium-medium scattering length is $k_na_\text{B} \simeq 0.01$. We, thus, assume that the relevant physics can be explained by assuming an ideal BEC. Here, the impurity self-energy is $\Sigma(\omega) = n_\text{B} \mathcal{T}(\omega)$, with the impurity-medium scattering matrix (evaluated for equal masses and at zero momentum)
\begin{equation}
\mathcal{T}(\omega) = \frac{\mathcal{T}_v}{1 - \mathcal{T}_v \Pi(\omega)}, 
\label{eq.T_matrix}
\end{equation}
where $\mathcal{T}_v = 4\pi \hslash^2 a / m$ is the zero-energy scattering matrix in vacuum, $\Pi(\omega) = -i m^{3/2}(\omega + i\eta)^{1/2}/(4\pi\hslash^{5/2})$ is the pair propagator, and $\eta$ is a positive infinitesimal. In turn, this yields the impurity Green's function (again at zero momentum)
\begin{equation}
G(\omega) = \frac{1}{\omega + i\eta - \Sigma(\omega)} = \frac{1}{\omega + i\eta - n_\text{B}\mathcal{T}(\omega)}, 
\end{equation}
from which the spectral function may also be obtained according to $A(\omega) = - 2{\rm Im} G(\omega)$. This is the so-called non-selfconsistent ladder approximation, from which a polaron energy at order $-E_n$ at strong coupling follows. It can readily be shown that the high-frequency tail equals the exact result in Eq. \eqref{eq.high_frequency_tail} for $a_\text{B} = 0$. However, this formalism also leads to an unphysical gap between the attractive polaron branch and the many-body continuum, starting at $\omega = 0$ according to Eq. \eqref{eq.T_matrix}. While this gap will likely vanish if a fully self-consistent approach is employed, this is in practice unfeasible to do. Instead, we take care of this in a phenomenological way by letting
\begin{equation}
\Pi(\omega) \to \Pi_R(\omega) = {\rm Re}[\Pi(\omega)] + i {\rm Im}[\Pi(\omega - E_\text{p}/\hslash)], 
\end{equation}
where the polaron energy is found by solving $E_\text{p}/\hslash = \Sigma(E_\text{p}/\hslash ) = n_\text{B}\mathcal{T}(E_\text{p}/\hslash )$. In this way, the polaron energy remains fixed at $E_\text{p}$, whereas the continuum now starts just at $\omega = E_\text{p} / \hslash$. One unfortunate consequence of this procedure is that the resulting spectral function, $A_R$, no longer obeys the correct normalization: $\int {\rm d}\omega A_R(\omega) / (2\pi) > 1$. To take care of this, the spectral function is renormalized. As a result, however, the high-frequency tail no longer perfectly matches the exact result, and therefore, the short-time dynamics is slightly altered. This effect is small for negative scattering lengths, $a < 0$, but becomes more pronounced on the positive side. \\

In the explicit computations, we take a small broadening of $\eta = 0.03 E_n$. To take care of the trap inhomogeneity, we use a local density approximation, in which the spectral function is simply averaged over the trap according to
\begin{equation}
\bar{A}(\omega) = \frac{1}{N} \int {\rm d}^3 r \; n_\text{B}(r) A(n_\text{B}(r),\omega), 
\label{eq.spectral_function_trap_average}
\end{equation}
where $n_\text{B}(r)$ is the local density of condensate atoms a distance $r$ from the center of the trap, evaluated within the Thomas-Fermi approximation, and $N$ is the total number of atoms. To obtain the impurity coherence function, we finally Fourier transform the trap-averaged spectral function Eq. \eqref{eq.spectral_function_trap_average}
\begin{equation}
C(t) = \int \frac{{\rm d}\omega}{2\pi} e^{-i(\omega + i\eta)t}\bar{A}(\omega).
\label{eq.coherence_function_trap_average}
\end{equation}
We have numerically checked that the resulting coherence function is unaltered, when the broadening is reduced by a factor of $10$.

\bibliography{ImpDynPhysRevRBib.bib}

\end{document}